\documentclass[11pt,bezier]{article}
\usepackage{amsmath,amssymb,amsfonts,euscript}
\usepackage{algorithmic, algorithm}
\usepackage{graphicx}

\usepackage[ansinew]{inputenc}
\usepackage{graphicx}
\usepackage{color}
\usepackage{mathrsfs}

\usepackage[colorlinks]{hyperref}
\usepackage[active,new,noold,marker]{xrcs}
\catcode`\=13
\def{$\bowtie$}

\usepackage{eurosym}
\usepackage{lscape} 

\newtheorem{prethm}{{\bf Theorem}}

\newenvironment{thm}{\begin{prethm}{\hspace{-0.5
               em}{\bf}}}{\end{prethm}}

\newtheorem{prepro}{{\bf Theorem}}

\newtheorem{preprop}{{\bf Proposition}}

\newtheorem{precor}{{\bf Corollary}}

\newtheorem{preconj}{{\bf Conjecture}}

\newtheorem{predefi}{{\bf Definition}}

\newtheorem{preexam}{{\bf Example}}

\newtheorem{preremark}{{\bf Remark}}

\newenvironment{remark}{\begin{preremark}\rm{\hspace{-0.5
               em}{\bf}}}{\end{preremark}}

\newtheorem{preexample}{{\bf Fact}}

\newtheorem{prelem}{{\bf Lemma}}

\newtheorem{prelam}{{\bf Lemma}}

\newtheorem{preprob}{{\bf Problem}}

\newenvironment{prob}{\begin{preprob}{\hspace{-0.5
               em}{\bf.}}}{\end{preprob}}

\newtheorem{preproof}{{\bf Proof}}

\newtheorem{preali}{{\bf Proof of Theorem 1.}}

\newenvironment{ali}[1]{\begin{preali}{\rm
               #1}\hfill{$\Box$}}{\end{preali}}

\newtheorem{prealii}{{\bf Proof of Theorem 2.}}

\newenvironment{alii}[1]{\begin{prealii}{\rm
               #1}\hfill{$\Box$}}{\end{prealii}}

\newtheorem{prealiii}{{\bf Proof of Theorem 3.}}

\newenvironment{aliii}[1]{\begin{prealiii}{\rm
               #1}\hfill{$\Box$}}{\end{prealiii}}

\newtheorem{prealiiii}{{\bf Proof of Theorem 4.}}

\newtheorem{prealij}{{\bf Proof of Theorem 5.}}

\newtheorem{prealijj}{{\bf Proof of Theorem 6.}}

\newtheorem{prealijjj}{{\bf Proof of Theorem 7.}}

\newtheorem{prealijjjk}{{\bf Proof of Theorem 8.}}

\title{On the semi-proper orientations of graphs}

\author{{\normalsize
		{ Ali Dehghan}\,
	}\vspace{3mm}
	\\{\footnotesize{\it Systems and Computer Engineering Department, Carleton University, Ottawa,   Canada}}
	\thanks{{\it E-mail address}:  $\mathsf{alidehghan@sce.carleton.ca}$. } }

\date{}
\begin{document}
\maketitle

\begin{abstract}
{\small \noindent
A  {\it semi-proper orientation} of a given graph $G$ is a function $(D,w)$ that assigns an orientation $D(e)$ and a positive integer weight $ w(e)$ to each edge $e$ such that for every two adjacent vertices $v$ and $u$, $S_{(D,w)}(v) \neq S_{(D,w)}(u) $, where  $S_{(D,w)}(v)  $ is the sum of the weights of edges with head $v$ in $D$.
The {\it semi-proper orientation number} of  a graph $G$, denoted by $\overrightarrow{\chi}_s (G)$, is $ \min_{(D,w)\in \Gamma}  \max_{v\in V(G)}  S_{(D,w)}(v)  $, where $\Gamma$ is the set of all semi-proper orientations of $G$. The {\it optimal semi-proper orientation} is a semi-proper orientation $(D,w)$ such that $ \max_{v\in V(G)} S_{(D,w)}(v)=  \overrightarrow{\chi}_s (G) $.
In this work, we show that every graph $G$ has an optimal semi-proper orientation $(D,w)$ such that the weight of each edge is one or two. Next, we show that determining whether a given planar graph $G$ with $\overrightarrow{\chi}_s (G)=2 $ has an optimal semi-proper orientation $(D,w)$ such that the weight of each edge is one is NP-complete. Finally, we prove that the problem of determining the semi-proper orientation number  of planar bipartite graphs is NP-hard. 
}

\begin{flushleft}

\end{flushleft}

\end{abstract}

\section{Introduction}
\label{}

A {\it directed graph} $G$ is an ordered pair $(V (G),E(G))$ consisting of a set
$V (G)$ of vertices and a set $E(G)$ of edges, with an incidence function $ D $ that associates with each edge of $G$ an ordered pair of vertices of $G$. If $e=uv$ is an edge and $D(e) = u \rightarrow v$, then
$e$ is from $u$ to $v$. The vertex $u$ is the tail of $e$, and the vertex $v$ is its head.
Let $G$ be an undirected graph  with no loop or parallel edges (i.e., simple graph). 
An orientation $D$ of a graph $G$ is a digraph obtained from the graph $G$ by replacing each edge by exactly one of the two possible arcs with the same endvertices. Also, the indegree $d_{D}^{-}(v)$ of a vertex $v$ in $D$ is the
number of edges with head $v$ in $D$.

\subsection{Proper Orientation}

An orientation of a graph $G$ is called {\it proper orientation} if any two adjacent vertices have different indegrees. The {\it proper
orientation number} of a graph $G$, denoted by $\overrightarrow{\chi} (G)$,  is the minimum of the maximum indegree taken over all proper orientations of the graph $G$. 
Note that the maximum indegree of vertices is $\Delta$. On the other hand,
the values of the indegrees define a proper vertex coloring of $G$ (adjacent vertices have different indegrees). Thus, 
\begin{equation}\label{E1}
 \chi(G)-1 \leq  \overrightarrow{\chi} (G)\leq \Delta(G)
\end{equation}

The existence of proper orientation was demonstrated by  Borowiecki,   Grytczuk and  Pil\'{s}niak in \cite{MR2895496}, where it was shown that every graph $G$ has a proper orientation $D$ with a vertex $v$ with $d_D^{-}(v)=\Delta$. 	
Afterwards, the proper orientation number   was introduced in  \cite{MR3095464}. Recently, the proper orientation has been studied extensively by several authors, for instance see \cite{MR3095464,  MR3704829,  MR3293286, MR3514378,   araujo2018weighted, MR3714524}. 

We should mention that interest in proper orientations stems from their connection to the 1-2-3-Conjecture that says: ``Can the edges of any non-trivial graph be assigned weights from $\{1,2,3\}$ so that adjacent vertices have different sums of incident edge weights?" \cite{MR2047539}. For more information about 1-2-3-Conjecture and its variants see \cite{MR2876224, MR3512668,  MR3268690,   MR3022926}.

\subsection{Semi-proper Orientation}

Motivated by the proper orientations of graphs and  1-2-3-Conjecture we investigate the semi-proper orientations of graphs.	
The {\it semi-proper orientation} of a given graph $G$ is a function $(D,w)$ that assigns an orientation $D(e)$ and a positive integer weight $ w(e)$ to each edge $e$ such that for every two adjacent vertices $v$ and $u$, $S_{(D,w)}(v) \neq S_{(D,w)}(u) $, where  $S_{(D,w)}(v)  $ is the sum of the weights of edges with head $v$ in $D$.
The {\it semi-proper orientation number} of  a graph $G$, denoted by $\overrightarrow{\chi}_s (G)$, is $ \min_{(D,w)\in \Gamma}  \max_{v\in V(G)}  S_{(D,w)}(v)  $, where $\Gamma$ is the set of all semi-proper orientations of $G$. 
Note that throughout the paper for each semi-proper orientation  $(D,w)$ of $G$ we denote $\sum_{z\in N_D^{-}(v)} w(zv)$  by $S_{(D,w)}(v)$.

Every proper orientation of a graph $G$ is a semi-proper orientation where the weights of all edges are one. Consequently, by (\ref{E1}), we have

\begin{equation}\label{E2}
\chi(G)-1 \leq  \overrightarrow{\chi}_s (G) \leq  \overrightarrow{\chi} (G)\leq \Delta(G)
\end{equation}

The {\it optimal semi-proper orientation} is a semi-proper orientation $(D,w)$ such that $ \max_{v\in V(G)} S_{(D,w)}(v)=  \overrightarrow{\chi}_s (G) $.
In this work, we show that every graph $G$ has an optimal semi-proper orientation $(D,w)$ such that for each edge $e\in E(G)$, we have $w(e)\in \{1,2\}$.

\begin{thm}\label{T1}
Every graph $G$ has an optimal semi-proper orientation $(D,w)$ such that the weight of each edge is one or two. 
\end{thm}

Although for each graph $G$ we can find an optimal semi-proper orientation $(D,w)$ such that the weight of each edge is one or two, there  are graphs such that they do not have any optimal  semi-proper orientation without edges with label two. 
In other words,  there are graphs such that for each of them $\overrightarrow{\chi}_s   <  \overrightarrow{\chi}  $. Next, we study the complexity of  finding those graphs.

\begin{thm}\label{T2}
It is  NP-complete to determine whether a given planar graph $G$ with $\overrightarrow{\chi}_s (G)=2 $ has an optimal semi-proper orientation $(D,w)$ such that the weight of each edge is one.
\end{thm}

\begin{remark}
Here, we show that for every tree $T$ we have $\overrightarrow{\chi}_s (T)\leq 2 $. Let $T$ be a tree and $v$ be a vertex in $T$. Run depth-first search (DFS) algorithm from the root $v$.  This defines a partition $\mathcal{L}_0, \mathcal{L}_1, \ldots, \mathcal{L}_h$ of the vertices of $T$ where each part $\mathcal{L}_i$ contains the vertices of $T$ which are at depth $i$ (i.e. at distance exactly $i$ from the vertex  $v$).  Note that by this partition each edge is between the vertices of to consecutive parts $\mathcal{L}_j$ and $\mathcal{L}_{j+1}$. Next, for each edge $e=u u'$, where $u \in \mathcal{L}_j$ and $u'\in\mathcal{L}_{j+1}$, orient $e$ from $u$ to $u'$. Call the resultant orientation $D$. In orientation $D$ the indegree of each vertex except the vertex $v$ is exactly one. (Also the indegree of $v$ is zero.) Finally, define the following weight function for the edges of $T$:
\begin{center}
	$w (e )=
	\begin{cases}
	1,    &   e =u u', \,\, u  \in \mathcal{L}_{2j},\,\, u'\in\mathcal{L}_{2j+1},\,\, j \geq 0,  \\
	2,       & \text{otherwise}.\
	\end{cases}$
\end{center}
It is easy to see that for each vertex $u\in \mathcal{L}_{2j}$, where $j \geq 1$, we have $S_{(D,w)}(u)=2$ and   for each vertex $u\in \mathcal{L}_{2i+1}$, where $i \geq 0$, we have $S_{(D,w)}(u)=1$ and also $S_{(D,w)}(v)=0$. Thus, $(D,w)$ is a semi-proper orientation of $T$ and we have $\overrightarrow{\chi}_s (T)\leq 2 $. This completes the proof.
\end{remark}

It was shown  \cite{MR3293286} that it is NP-complete to decide whether the proper orientation number of a given planar bipartite graph  is less than or equal to three. We improve this hardness result for semi-proper orientation number and proper orientation number of graphs.

\begin{thm}\label{T3}
(1) For a given planar bipartite graph $G$ determining whether  $\overrightarrow{\chi}_s (G)\leq 2 $ is NP-complete.\\
(2) For a given planar bipartite graph $G$ determining whether  $\overrightarrow{\chi}  (G)\leq 2 $ is NP-complete.
\end{thm}

The organization of the rest of the paper is as follows: In Section \ref{S2}, we present some definitions and notations. Next, in Section \ref{S3}, we prove that   every graph $G$ has an optimal semi-proper orientation $(D,w)$ such that the weight of each edge is one or two.
In Section \ref{S4}, we study the computational complexity of finding optimal semi-proper orientations and computing the semi-proper orientation number of planar bipartite graphs. The paper is concluded with some remarks and open problems in Section \ref{S5}

\section{Notation}\label{S2}

Throughout  this paper we only consider finite simple  graphs (i.e. finite graphs with no loop or parallel edges). We denote the vertex set and the edge set of $G$ by $V(G)$ and $E(G)$,
respectively. Also, we denote the maximum degree and the minimum degree
of $G$ by $\Delta(G)$ and $\delta(G)$, respectively. 

A proper vertex coloring of a graph $G$ is a function $ f:V(G) \rightarrow L$ such that if $v,u \in V(G)$ are adjacent, then $f(u)$ and $f(v)$ are
different. A proper vertex $k$-coloring is a proper vertex coloring with $|L| = k$. The smallest integer $k$ such that the graph $G$ has a proper
vertex $k$-coloring is called the chromatic number of $G$ and denoted by $\chi(G)$.

Consider the graph $G = (V,E)$, and let $U \subset V$ be any
subset of vertices of $G$. Then, the  induced subgraph   on the set of vertices $U$ is the graph
whose vertex set is $U$ and whose edge set consists of all the
edges in $E$ that have both endpoints in $U$.
We follow \cite{MR1367739} for terminology and notation
where they are not defined here.

\section{1-2-result for semi-proper orientations}\label{S3}

In this section we prove that every graph $G$ has an optimal semi-proper orientation $(D,w)$ such that the weight of each edge is one or two. 

\begin{ali}{ 
We prove the theorem by using contradiction. To the contrary suppose that $G$ is a graph such that each of its optimal semi-proper orientations has an edge with weight more than two. In an optimal semi-proper orientation of a graph $G$ if the weight of an edge $e$ is more than two then we say that $e$ has a {\it bad label} in that semi-proper orientation. Without loss of generality suppose that each  optimal
semi-proper orientations of $G$ has at least $t$, $t>0$, edges with bad label and there is an optimal semi--proper orientation  such that it has exactly $t$ edges with bad label. Among all optimal
semi-proper orientations of $G$, let $ \mathcal{F}$ be the set of  optimal semi-proper orientations such that each of them has exactly $t$ edges with bad label. Clearly, $\mathcal{F}$ is non-empty. Finally, among the optimal
semi-proper orientations in $ \mathcal{F}$, let $(D,w)$ be an optimal semi-proper orientation   such that the sum of the weights of edges with bad labels  is minimum.
We assume that
the optimal semi-proper orientation $ (D,w)$ of $G$ has $t$ edges with bad label and the sum of the weights of  edges with bad label is $ b$.

Let $e=vu$  be an   edge with bad label $\alpha$ in $G$ (i.e. $w(uv)=\alpha$) and $D(e) = u \rightarrow v$. Also, let $\mathcal{R}_v$ be the set of vertices such that for each of them there is directed path from the vertex $v$ to that vertex in $D$. Among all vertices in $\mathcal{R}_v$ let $p$ be a vertex such that the sum of weights of incoming edges to $v$ is minimum (i.e. $  S_{(D,w)}(p) $ is minimum over all $\{S_{(D,w)}(z) | z\in \mathcal{R}_v\}$). 
Two cases for the vertex $p$ can be considered.
\\ \\
{\bf \underline{Case 1.}} Assume that $ S_{(D,w)}(p)= S_{(D,w)}(v)$. In this case the sum of weights of incoming edges to $v$ is minimum over all vertices in $\mathcal{R}_v$.  Let $ u_1,\ldots , u_k$ be the set of vertices such that from  each of them there is a directed edge to $v$. Without loss of generality assume that $u=u_1$. Also, let $ o_1,\ldots, o_r$ be the set of vertices such that for each of them there is   a directed edge from $v$ to that vertex.
So, we have $ o_1,\ldots, o_r \in \mathcal{R}_v $. Now, we study the properties of the vertex $v$.
\\ \\
{\bf Property 1.} For each $o_i\in \{o_1,\ldots , o_r\}$, we have $ S_{(D,w)}(v) <  S_{(D,w)}(o_i)$. 
\\ \\
{\bf Proof of Property 1.} By our assumption for each $o_i\in \{o_1,\ldots , o_r\}$, we have $ S_{(D,w)}(v) \leq   S_{(D,w)}(o_i)$. Since $S_{(D,w)}$ is a proper vertex coloring, for each $o_i\in \{o_1,\ldots , o_r\}$, we have $ S_{(D,w)}(v) <  S_{(D,w)}(o_i)$.  $\blacksquare$
\\ \\
{\bf Property 2.} For each edge $u_i v$, $2 \leq i \leq k$,   we have $ w(u_iv)=2$. 
\\ \\
{\bf Proof of Property 2.} To the contrary assume that there is a vertex $u_l$, $2 \leq l \leq k$, such that $ w(u_lv)\neq 2$. Two situations can be considered. \\
$ \bullet$ If $ w(u_lv)= 1$, then consider the  function $(D',w')$, where
\begin{center}
	$w'(e')=
	\begin{cases}
	 w(uv)-1,    &   e'=uv,   \\
	 w(u_l v)+1,    &  e'=u_l v ,      \\
     w(e'),       & \text{otherwise},\
	\end{cases}$
\end{center}
and $D'=D$. Clearly,  $(D',w')$
is an optimal semi-proper orientation such that  it has at most $t$ edges with bad label and the sum of the weights of  edges with bad label is less than $ b$. But this is a contradiction. \\
$ \bullet$ If $ w(u_lv)> 2$,  
 then consider the  function $(D',w')$, where
\begin{center}
	$w'(e')=
	\begin{cases}
	1,    &   e'=uv ,  \\
	w(u_l v)+w(uv)-1,    &  e'=u_l v,       \\
	w (e'),       & \text{otherwise},\
	\end{cases}$
\end{center}
and $D'=D$. Clearly,  $(D',w')$ is an optimal semi-proper orientation such that  it has   $t-1$ edges with bad label. But this is a contradiction. This completes the proof of the Property 2. $\blacksquare$ 
\\ \\
{\bf Property 3.} For each $o_i\in \{o_1,\ldots , o_r\}$, we have $  S_{(D,w)}(o_i)> 2k+1$. 
\\ \\
{\bf Proof of Property 3.} We have $w(u_1 v)=\alpha > 2$, so by Property 2, we have $  S_{(D,w)}(v)> 2k$. Thus, by Property 1, we have $  S_{(D,w)}(o_i)> 2k+1$. $\blacksquare$

Now, we are ready to prove this case. For each edge $u_i v$, $1 \leq i \leq k$, define a variable $Var(u_iv)$. Each variable can be one or two. The sum of variables  is an integer between $k$ and $2k$. 
(i.e. $  k \leq \sum_i Var(u_iv) \leq 2k$). So the sum of the variables can be $k+1$ different integers. Consequently, there is an integer $q$ such that 
$k \leq q \leq 2k$ and $q \notin \{ S_{(D,w)}(u_i): 1 \leq i \leq k\}$. Assign one and two to the variables such that their sum is $q$. Now, define the  function $(D',w')$, where
\begin{center}
	$w'(e')=
	\begin{cases}
	Var(u_iv),    &   e'=u_i v,\, \, 1 \leq i \leq k ,  \\
	w (e'),       & \text{otherwise},\
	\end{cases}$
\end{center}
and $D'=D$. By Property 3, and the way that we choose $q$, it is clear that $(D',w')$ is an optimal semi-proper orientation such that  it has   $t-1$ edges with bad label. But this is a contradiction.
\\ \\
{\bf \underline{Case 2.}} Assume that $S_{(D,w)}(p) \neq S_{(D,w)}(v)$. By the definition of 
$\mathcal{R}_v$ there is a directed path $\mathcal{P}=v,z_1,\ldots, z_l,p$ from the vertex $v$ to the vertex $p$.  Let $ u_1,\ldots , u_k$ be the set of vertices such that from each of them there is  a directed edge to $p$. Without loss of generality assume that $z_l=u_1$.  (Note that we can have the case where the only directed path form $v$  to $p$ is the edge $\overrightarrow{vp}$. In that case we assume that $u_1=v$.) Next, let $ o_1,\ldots, o_r$ be the set of vertices such that for each of them there is a directed edge from $p$ to that vertex. Now, we study the properties of the vertex $p$. 
\\ \\
{\bf Property 4.} We have $  S_{(D,w)}(p)\geq  w(u_1p)+ 2k-3 $. 
\\ \\
{\bf Proof of Property 4.} To the contrary suppose that $  S_{(D,w)}(p) \leq  w(u_1p)+ 2k-4 $. So there are indexes $j,j'$, where $j,j'\neq 1$  and $ w(u_jp)= w(u_{j'} p)=1$. 
First, we define a new notation and then we complete the proof. Let $z_1,z_2$ be two arbitrary adjacent vertices (i.e. $e'=z_1z_2 \in E(G)$). If  we have $D(e')=z_1 \rightarrow z_2$, then we denote $z_2 \rightarrow z_1$ by $ \neg D(e')  $.
\\
Now, define the  function $(D',w')$, where
\begin{center}
	$D'(e')=
	\begin{cases}
	\neg D(e'),   &  e'\in \mathcal{P},\\
	D(e'),       & \text{otherwise},\
	\end{cases}$
\end{center}
and
\begin{center}
	$w'(e')=
	\begin{cases}
	\alpha-1 ,          &   e'=uv, \\
	1  ,                &   e'=z_1 v,\\
	w(vz_1) ,           &   e'=z_2z_1, \\
	w(z_{i-1}z_{i}),    &   e'=z_{i+1} z_{i},\, 2 \leq i < l,\\
	w(z_{l-1}z_l),      &   e'= p z_l, \\
	2,                 &   e' =u_jp,  \\
	w(u_1p) ,           &   e' = u_{j'}p\\
	w (e'),            & \text{otherwise}.\
	\end{cases}$
\end{center}
Note that for every vertex $ f$ we have $ S_{(D,w)}(f)=  S_{(D',w')}(f)$. Thus,  $(D',w')$
is an optimal semi-proper orientation such that  it has at most $t$ edges with bad label and the sum of the weights of  edges with bad label is $ b-1$. But this is a contradiction. $\blacksquare$
\\ \\
{\bf Property 5.} For each $o_i\in \{o_1,\ldots , o_r\}$, we have $  S_{(D,w)}(o_i)>   2k-2$.
Also, we have $    S_{(D,w)}(z_l)>   2k-2 $. 
\\ \\
{\bf Proof of Property 5.} By our assumption  the sum of weights of incoming edges to the vertex  $p$ is minimum. Thus, for each $o_i\in \{o_1,\ldots , o_r\}$, we have $ S_{(D,w)}(p) \leq   S_{(D,w)}(o_i)$. Also, $ S_{(D,w)}(p) \leq   S_{(D,w)}(z_l)$.  On the other hand, the function $S_{(D,w)}$ is a proper vertex coloring, so by Property 4, for each $o_i\in \{o_1,\ldots , o_r\}$, we have $  2k-2 < S_{(D,w)}(o_i)$. Also, we have $ 2k-2 <   S_{(D,w)}(z_l)$. $\blacksquare$

Now, we are ready to prove Case 2. For each edge $u_i p$, $2 \leq i \leq k$, define a variable $Var(u_ip)$. Each variable can be one or two. The sum of variables  is an integer between $k-1$ and $2k-2$. 
(i.e. $  k-1 \leq \sum_{ i=2}^{k} Var(u_ip) \leq 2k-2$). 
So the sum of the variables can be $k $ different
integers.
Hence, there is an integer $q$ such that 
$k-1 \leq q \leq 2k-2$ and $q \notin \{ S_{(D,w)}(u_i): 2 \leq i \leq k\}$. Assign one and two to the variables such that their sum is $q$. Now, define the  function $(D',w')$, where
\begin{center}
	$D'(e')=
	\begin{cases}
	\neg D(e'),   &  e'\in \mathcal{P},\\
	D(e'),       & \text{otherwise},\
	\end{cases}$
\end{center}
and
\begin{center}
	$w'(e')=
	\begin{cases}
	\alpha-1 ,          &   e'=uv, \\
	1  ,                &   e'=z_1 v,\\
	w(vz_1) ,           &   e'=z_2 z_1, \\
	w(z_{i-1}z_{i}),    &   e'= z_{i+1} z_{i},\, 2 \leq i < l,\\
	w(z_{l-1}z_l),      &   e'= p z_l, \\
	Var(u_iv),           &   e'=u_i v,\, \, 2 \leq i \leq k ,  \\
	w (e'),            & \text{otherwise}.\
	\end{cases}$
\end{center}
By Property 5, and the way that we choose $q$, it is clear that $(D',w')$ is an optimal semi-proper orientation such that  it has at most  $t$ edges with bad label and the sum of the weights of  edges with bad label is at most $ b-1$. But this is a contradiction. This completes the proof.
}\end{ali}

\section{Hardness results}\label{S4}

First, we introduce planar $3$-SAT (type $2$) formula.
Let $\Phi$ be a $3$-SAT formula with the set of clauses $C =\{c_1, \cdots ,c_k\}$ and the set of variables
$X=\lbrace x_1, \cdots ,x_n\rbrace $. Let $\mathcal{G}_\Phi$ be a graph with the set of  vertices $C \cup X \cup (\neg X)$, where $\neg X = \lbrace \neg x_1, \cdots , \neg x_n\rbrace$, such that for each clause $c_j=(y \vee z \vee w) $, the vertex $c_j$ is adjacent to the vertices $y$, $z$ and $w$. Also every vertex $x_i \in X$ is adjacent to the vertex $\neg x_i$. The formula $\Phi$ is called planar $3$-SAT (type $2$)   if the graph $\mathcal{G}_\Phi$ is a planar graph. Throughout the paper we refer to $\mathcal{G}_\Phi$ as the type 2 graph that was derived from the formula $\Phi$.
It was  proved  that the problem of determining the satisfiability of planar $3$-SAT (type $2$) is $ \mathbf{NP}$-complete \cite{zhu1}.
\\ \\
\textbf{Problem}: {\em Satisfiability of planar $3$-SAT (type $2$).}\\
\textsc{Input}: A planar $3$-SAT (type $2$) formula  $ \Phi $.\\
\textsc{Question}: Is there a truth assignment for $ \Phi $ that satisfies all the clauses?\\ 
 
Next, by using a polynomial time reduction from  {\em Satisfiability of planar $3$-SAT (type $2$)}, we show that it is NP-complete to determine whether a given planar graph $G$ with $\overrightarrow{\chi}_s (G)=2 $ has an optimal semi-proper orientation $(D,w)$ such that the weight of each edge is one. 

\begin{alii}{

Let $\Phi $ be an instance of planar $3$-SAT(type $2$) formula 		
 with the set of  variables
$X=\lbrace x_1, \cdots ,x_n\rbrace $ and the set of clauses $C=\lbrace
c_1, \cdots ,c_k\rbrace $. We transform this formula into a planar graph $\mathcal{H}_\Phi$ such that $\overrightarrow{\chi}_s (\mathcal{H}_\Phi)=2 $ and the graph $\mathcal{H}_\Phi$ has an optimal semi-proper orientation $(D,w)$ such that the weight of each edge is one if and only if there is a satisfying assignment for $\Phi $.  First, we introduce two useful gadgets.
\\ \\
{\bf Property 6.} Consider the gadget $\mathcal{T}_{x_i}$ which is shown in Fig. \ref{F2}. Let    $(D,w)$ be an optimal semi-proper orientation of  $\mathcal{T}_{x_i}$ such that for every edge $e$ we have $w(e)=1$. Then $\{S_{(D,w)}(x_i), S_{(D,w)} (\neg x_i)\}=  \{1,2\}$.
\\ \\
{\bf Proof of Property 6.} It is easy to see that $\overrightarrow{\chi}_s (\mathcal{T}_{x_i})=2 $.  Let $(D,w) $ be an optimal semi-proper orientation of $\mathcal{T}_{x_i}$ such that for every edge $e$ we have $w(e)=1$. The induced subgraph on the set of three vertices $\{p_2, x_i, \neg x_i\}$ forms a cycle of length three. So,  $$\{S_{(D,w)}(p_2),S_{(D,w)}(x_i), S_{(D,w)} (\neg x_i)\}=  \{0,1,2\}.$$ Thus, the two edges $p_2p_1$ and $p_2p_3$ were oriented from $p_2$ to $p_1$ and $p_3$, respectively. (Note that  for every edge $e$ we have $w(e)=1$.) Therefore, $S_{(D,w)}(p_1)=1$ and $S_{(D,w)}(p_3)\in \{1,2\}$. Next, we show that $S_{(D,w)}(p_3)=2$. To the contrary suppose that $S_{(D,w)}(p_3)=1$. So, the edges  $p_3p_4$ and $p_3p_5$ were oriented from $p_3$ to $p_4$ and $p_5$, respectively. Thus, $S_{(D,w)}(p_4)=S_{(D,w)}(p_5)=1$. So for two adjacent vertices $p_3$ and $p_4$, we have $S_{(D,w)}(p_3)=S_{(D,w)}(p_4)$. But this is a contradiction. So $S_{(D,w)}(p_3)=2$. Consequently, $\{S_{(D,w)}(p_1),S_{(D,w)}(p_3) \}=\{1,2\}$. Thus, $S_{(D,w)}(p_2)=0$. Hence
$\{S_{(D,w)}(x_i), S_{(D,w)} (\neg x_i)\}=  \{1,2\}$. $\blacksquare $

\begin{figure}[ht]
	\begin{center}
		\includegraphics[scale=.40]{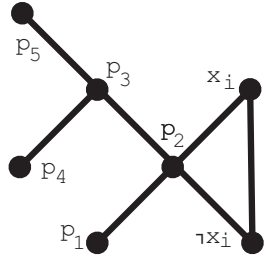}
		\caption{The gadget $\mathcal{T}_{x_i}$.} \label{F2}
	\end{center}
\end{figure}

In our proof we also use the gadget $\mathcal{K}_{c_j}$ which is shown in Fig. \ref{F3}. Next, we present the reduction.\\
\underline{Construction of $\mathcal{H}_\Phi$.} 
We construct the planar graph $\mathcal{H}_\Phi$ from the   type 2 graph $ \mathcal{G}_\Phi $ (that was derived from the formula $\Phi$) in two steps. \\
{\bf Step 1.} For each variable $x_i \in X$ put a copy of the gadget $\mathcal{T}_{x_i}$. We call these {\it variable gadgets}. Also, for  every clause  $c_j \in C$ put a copy of the gadget $\mathcal{K}_{c_j}$.   We call these {\it clause gadgets}.\\ 
{\bf Step 2.} For every clause $c_j\in C$ with the literals $x$, $y$, $z$ (i.e. $c_j=(x \vee y \vee z)$, where $x,y,z\in X \cup (\neg X)$) add the edges $xc^j_1$, $yc^j_2$ and $zc^j_3$.
Call the resultant graph $\mathcal{H}_\Phi$.

\begin{figure}[ht]
	\begin{center}
		\includegraphics[scale=.40]{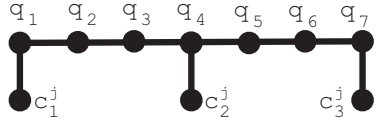}
		\caption{The gadget $\mathcal{K}_{c_j}$.} \label{F3}
	\end{center}
\end{figure}

Clearly, the resultant graph is planar and   we have $\overrightarrow{\chi}_s (\mathcal{H}_\Phi)\geq 2 $. Next, we study an important property of  $\mathcal{K}_{c_j}$.
\\ \\
{\bf Property 7.}
Let $\overrightarrow{\chi}_s (\mathcal{H}_\Phi)= 2 $ and $(D,w)$ be an optimal  semi-proper orientation of $\mathcal{H}_\Phi$ such that for every edge $e$ we have $w(e)=1$. Then every edge  that connects a vertex of a clause gadget  to a vertex of a variable gadget  is oriented from   the vertex of the variable gadget  to  the vertex of the clause gadget.
\\ \\
{\bf Proof of Property 7.} Let $e= x_i c_1^j$ be one of those edges. The induced graph on the set of three vertices $\{p_2, x_i, \neg x_i\}$ forms a cycle of length three. Thus,  $$\{S_{(D,w)}(p_2),S_{(D,w)}(x_i), S_{(D,w)} (\neg x_i)\}=  \{0,1,2\}.$$  So, the  edge  $x_i c_1^j$ was oriented from $x_i$ to $c_1^j $. $\blacksquare $
\\ \\
{\bf Property 8.} Consider the graph $\mathcal{H}_\Phi$. Let $c_j \in C $ be an arbitrary clause and $(D,w)$ be an optimal  semi-proper orientation of $\mathcal{H}_\Phi$ such that for every edge $e$ we have $w(e)=1$. Also, assume that  $\overrightarrow{\chi}_s (\mathcal{H}_\Phi)= 2 $. Then   $2\in \{S_{(D,w)}(c_1^j), S_{(D,w)} (c_2^j) , S_{(D,w)} (c_3^j)\}    $.
\\ \\
{\bf Proof of Property 8.} Consider the subgraph  $\mathcal{K}_{c_j}$. To the contrary suppose that $S_{(D,w)}(c_1^j)= S_{(D,w)} (c_2^j) = S_{(D,w)} (c_3^j)=1 $. By property 7, and our assumption about the values of the function $S_{(D,w)}$ for the vertices $c^j_1 ,c^j_2 ,c^j_3 $, we can conclude that the edges $c^j_1 q_1 $, $c^j_2 q_4 $ and $c^j_3 q_7 $ should be oriented form $ c^j_1$, ($c^j_2 ,c^j_3 $, respectively) to $q_1$, ($q_4,q_7$, respectively). Thus, we have $S_{(D,w)}(q_1)= S_{(D,w)} (q_4) = S_{(D,w)} (q_7)=2 $. Since $(D,w)$ is a proper orientation we have $\{ S_{(D,w)}(q_2), S_{(D,w)} (q_3) \} = \{ 0,1\} $ and $\{ S_{(D,w)}(q_5), S_{(D,w)} (q_6) \} = \{ 0,1\} $. Thus the two edges $q_3q_4, q_5q_4$ were oriented from $q_3$, ($q_5$, respectively) to the vertex $q_4$. But this shows that the incoming degree of the vertex $q_4$ is three which is a contradiction. $\blacksquare $

\begin{figure}[ht]
	\begin{center}
		\includegraphics[scale=.40]{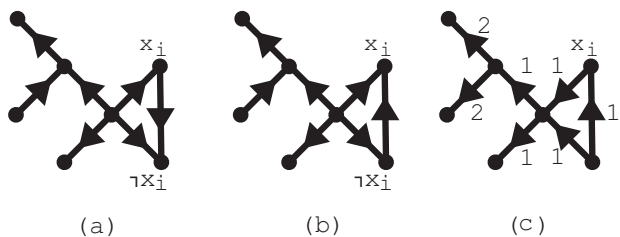}
		\caption{Three useful ways to assign orientation and weight to the edges of the gadget $\mathcal{T}_{x_i}$. Note that in (a) and (b) the weight of any edge is one.} \label{F4}
	\end{center}
\end{figure}

Next, we show that the graph $\mathcal{H}_\Phi$ has an optimal semi-proper orientation $(D,w)$ such that the weight of each edge is one if and only if there is a satisfying assignment for $\Phi $.\\ \\
\underline{Proof of $\Rightarrow$}. Let $(D,w)$ be an  optimal semi-proper orientation such that the weight of each edge is one. Let $\Gamma : X \rightarrow \{ {\sf{\sf true}},{\sf false} \} $ be a function such that if $  S_{(D,w)}(x_i)=1 $, then $\Gamma( x_i)= {\sf{\sf true}}$ and if $  S_{(D,w)} (x_i)=2 $, then $\Gamma( x_i)= {\sf false}$. Next, we prove that $\Gamma$  is a satisfying assignment for $\Phi$. For each clause $c_j=(x \vee y \vee z)$, by Property 8, we have $2\in \{S_{(D,w)}(c_1^j), S_{(D,w)} (c_2^j) , S_{(D,w)} (c_3^j)\}    $. Thus, by Properties 6 and 7,  we have $1\in \{S_{(D,w)}(x), S_{(D,w)} (y) , S_{(D,w)} (z)\}    $. Consequently, the function $\Gamma$  is a satisfying assignment for $\Phi$.\\ \\
\underline{Proof of $\Leftarrow$}. Assume that $ \Phi $ is satisfiable with the satisfying
assignment $\Gamma : X \rightarrow \{ {\sf{\sf true}},{\sf false} \} $. Next, in three steps, we present an optimal semi-proper orientation such that the weight of each edge is one. \\
$ \bullet$ If $e$ is an edge that connects a vertex of a clause gadget  to a vertex of a variable gadget, then  orient $e$ from   the vertex of the variable gadget  to  the vertex of the clause gadget.\\
$ \bullet$ For each gadget $\mathcal{T}_{x_i}$ if $\Gamma(x)= {\sf{\sf true}}$, then orient the edges of $\mathcal{T}_{x_i}$ like Fig. \ref{F4}.(a). Otherwise orient those edges  like Fig. \ref{F4}.(b). \\
$ \bullet$ For each gadget $\mathcal{K}_{c_j}$, where $c_j=(x \vee y \vee z)$, we have 
 $1\in \{S_{(D,w)}(x), S_{(D,w)} (y) , S_{(D,w)} (z)\}    $. Consequently, we have $2\in \{S_{(D,w)}(c_1^j), S_{(D,w)} (c_2^j) , S_{(D,w)} (c_3^j)\}    $. Thus base on the values of $S_{(D,w)}(c_1^j), S_{(D,w)} (c_2^j) , S_{(D,w)} (c_3^j)$ use one of the orientations that were presented in Fig. \ref{F5}. This completes the proof of $(\Leftarrow)$.

\begin{figure}[ht]
	\begin{center}
		\includegraphics[scale=.40]{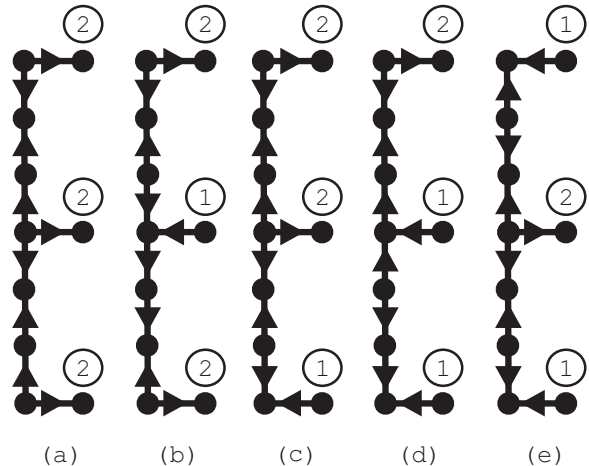}
		\caption{Possible ways to assign orientations   to the edges of the gadget $\mathcal{K}_{c_j}$.  } \label{F5}
	\end{center}
\end{figure}

Finally, we show that the semi-proper orientation number of the graph $\mathcal{H}_\Phi$ is always two. We present an optimal semi-proper orientation $(D,w)$ such that the weights of some edges are two. Consider the following orientations and weights for the edges of the graph.\\
$ \bullet$ If $e$ is an edge that connects a vertex of a clause gadget  to a vertex of a variable gadget, then put $w(e)=1$ and  orient $e$ from   the vertex of the variable gadget  to  the vertex of the  clause gadget.\\
$ \bullet$ For each gadget $\mathcal{T}_{x_i}$ put orientations and weights on  the edges of $\mathcal{T}_{x_i}$ like Fig. \ref{F4}.(c)\\
$ \bullet$ For each gadget $\mathcal{K}_{c_j}$, put the weight of any edge one and orient them like Fig. \ref{F5}.(a). This completes the proof.

}\end{alii}

In Part (1) of the next proof we show that for a given planar bipartite graph $G$ determining whether  $\overrightarrow{\chi}_s (G)\leq 2 $ is NP-complete. After that in Part (2), we show that  computing   $\overrightarrow{\chi}_s (G) $ for  planar bipartite graphs is NP-hard.

\begin{aliii}{ 
(1) It is clear that the problem is in NP. We reduce {\em  Cubic planar 1-in-3 SAT} to our problem in polynomial time. 
\\ \\
{\em  Cubic planar 1-in-3 SAT.}\\
\textsc{Instance}: A 3-SAT formula $\Phi=(X,C)$
such that every variable
appears in exactly three clauses, there
is no negation in the formula, and the
bipartite graph obtained by linking a variable and a clause if and only
if the
variable appears in the clause, is planar.\\
\textsc{Question}: Is there a truth assignment for $X$ such that
each clause in $C$ has exactly
one {\sf{\sf true}} literal?

In 2001, Moore and Robson proved
that {\em  Cubic planar 1-in-3 SAT} is NP-complete  \cite{MR1863810}.

\begin{figure}[ht]
	\begin{center}
		\includegraphics[scale=.30]{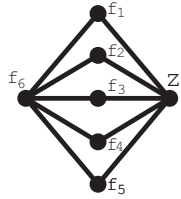}
		\caption{The gadget $\mathcal{F}_1$. } \label{F7}
	\end{center}
\end{figure}

Consider an instance $ \varphi $ of {\em  Cubic planar 1-in-3 SAT}. We transform this into a  planar bipartite graph $G$
such that  the semi-proper orientation number of $G$ is two
if and only if the formula  $ \varphi $ has a 1-in-3  assignment. First, we introduce some useful gadgets.
\\ \\
{\bf Property 9.} Let $G$ be graph such that $\overrightarrow{\chi}_s (G)=2 $ and it has the  gadget $\mathcal{F}_1$ (which is shown in Fig. \ref{F7}) as an induced subgraph.
Also, let $(D,w)$ be an optimal semi-proper orientation of $G$. Then $S_{(D,w)}(z)=\{0,1\} $.
\\ \\
{\bf Proof of Property 9.} To the contrary suppose that $S_{(D,w)}(z)=2 $. So by  the symmetry of the gadget without loss of generality we can assume that the edges $zf_1,zf_2,zf_3$ were oriented from $z$ to $f_1$ ($f_2$ and $f_3$, respectively). Since $S_{(D,w)}$ is a proper vertex coloring, we have $ S_{(D,w)}(f_1)= S_{(D,w)}(f_2)= S_{(D,w)}(f_3)=1$. Thus, the three edges $f_1f_6, f_2f_6, f_3f_6$ were oriented form $f_1$ ($f_2$, $f_3$, respectively) to $f_6$. Hence $S_{(D,w)}(f_6)\geq 3 $. But this is a contradiction. This completes the proof. $\blacksquare$
\\ \\
{\bf Property 10.} Let $G$ be graph such that $\overrightarrow{\chi}_s (G)=2 $ and  it has the  gadget $\mathcal{F}_2$ (which is shown in Fig. \ref{F8}) as an induced subgraph.
Also, let $(D,w)$ be an optimal semi-proper orientation of $G$. Then $S_{(D,w)}(z_2)=0 $.
\\ \\
{\bf Proof of Property 10.} By Property 9, we have $S_{(D,w)}(z_1), S_{(D,w)}(z_2), S_{(D,w)}(z_3)\in\{0,1\} $. To the contrary suppose that $S_{(D,w)}(z_2)=1 $. Then  at least one of the edges $z_1z_2, z_2z_3$ were oriented form $z_2$ to the other endpoint. Without loss of generality assume that  $z_2z_3$ were oriented form $z_2$ to $z_3$. So, $S_{(D,w)}(z_2)= S_{(D,w)}(z_3)=1$. But this is a contradiction.   $\blacksquare$

\begin{figure}[ht]
	\begin{center}
		\includegraphics[scale=.30]{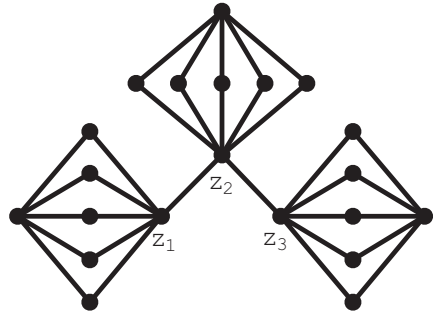}
		\caption{The gadget $\mathcal{F}_2$. } \label{F8}
	\end{center}
\end{figure}
 
{\bf Property 11.} Let $G$ be graph such that $\overrightarrow{\chi}_s (G)=2 $ and it has the variable gadget $\mathcal{H}_{x_i}$ (which is shown in Fig. \ref{F9}) as an induced subgraph.
Also, let $(D,w)$ be an optimal semi-proper orientation of $G$. Then $\{S_{(D,w)}(x_i),S_{(D,w)}(\neg x_i)\} =\{1,2\} $. 
\\ \\
{\bf Proof of Property 11.} By Property 10, we have $S_{(D,w)}(z_2)= S_{(D,w)}(z_2')=0$. Thus, the edges $z_2 x_i$ and $z_2' \neg x_i$ were oriented from $z_2$ ($z_2'$, respectively) to $x_i$ ($\neg x_i$, respectively). Consequently, $\{S_{(D,w)}(x_i),S_{(D,w)}(\neg x_i)\} =\{1,2\} $.  $\blacksquare$

\begin{figure}[ht]
	\begin{center}
		\includegraphics[scale=.35]{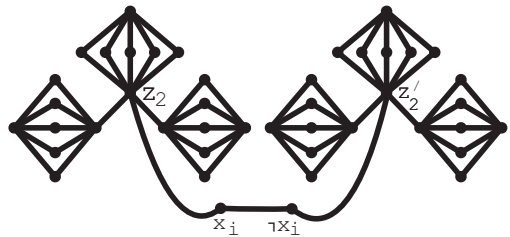}
		\caption{The gadget $\mathcal{H}_{x_i}$. } \label{F9}
	\end{center}
\end{figure}
 
{\bf Property 12.} Let $G$ be graph such that  $\overrightarrow{\chi}_s (G)=2 $ and it has the   gadget $\mathcal{Q}_{c_j}$ (which is shown in Fig. \ref{F10}) as an induced subgraph.
Also, let $(D,w)$ be an optimal semi-proper orientation of $G$. If there are two edges $xc_1^j$ and $yc_2^j$ such that $D(xc_1^j) = x \rightarrow c_1^j $ and $D(yc_2^j) = y \rightarrow c_2^j $. Then $2 \in \{S_{(D,w)}(c_1^j),S_{(D,w)}(c_2^j)\}   $. 
\\ \\
{\bf Proof of Property 12.} By Properties 9 and 10, we have $S_{(D,w)}(z_1)=1$. On the other hand, by Property 8, we should have $2\in \{ S_{(D,w)}(z_1),S_{(D,w)}(c_1^j), S_{(D,w)}(c_2^j)\}$. (Note that although in Property 8, we assume that the weight of any edge is one, the proof for the case where the weights are in $\{1,2\}$ is still correct.) Thus, we have  $2 \in \{S_{(D,w)}(c_1^j),S_{(D,w)}(c_2^j)\}   $. $\blacksquare$

\begin{figure}[ht]
	\begin{center}
		\includegraphics[scale=.30]{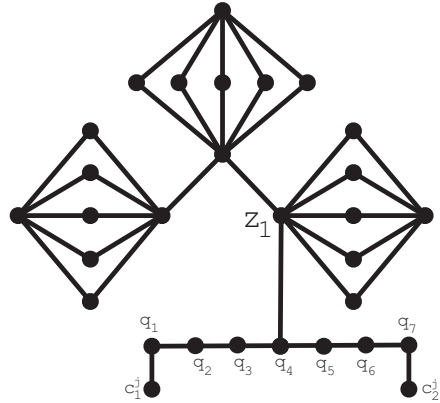}
		\caption{The gadget $\mathcal{Q}_{c_j}$. } \label{F10}
	\end{center}
\end{figure}

{\bf Property 13.} Let $G$ be graph such that $\overrightarrow{\chi}_s (G)=2 $ and it has the   gadget $\mathcal{L}_{c_j}$ (which is shown in Fig. \ref{F11}) as an induced subgraph.
Also, let $(D,w)$ be an optimal semi-proper orientation of $G$. Then  $   \{S_{(D,w)}(c_1^j),S_{(D,w)}(c_2^j) ,S_{(D,w)}(c_3^j)\}  \neq \{2,2,2\} $. 
\\ \\
{\bf Proof of Property 13.} To the contrary assume that $   \{S_{(D,w)}(c_1^j),S_{(D,w)}(c_2^j),S_{(D,w)}(c_3^j)\} = \{2,2,2\} $.  The semi-proper orientation number of $G$ is two, so at least one of the three edges $ p q_2, p q_4, pq_6 $ was oriented from $ p$ to the other endpoint. By the symmetry suppose that $ p q_2$ was oriented from $p$ to $ q_2$. On the other hand, by property 9, we have $S_{(D,w)}(q_2)  \in \{0,1\} $. Thus, $S_{(D,w)}(q_2)  =1 $. So, the edge  $q_2q_1$ was oriented from $q_2$ to $q_1$. Thus, $S_{(D,w)}(q_1)  =2 $, but this is a contradiction with $   \{S_{(D,w)}(c_1^j),S_{(D,w)}(c_2^j) ,S_{(D,w)}(c_3^j)\}  = \{2,2,2\} $. This completes the proof.  $\blacksquare$

\begin{figure}[ht]
	\begin{center}
		\includegraphics[scale=.35]{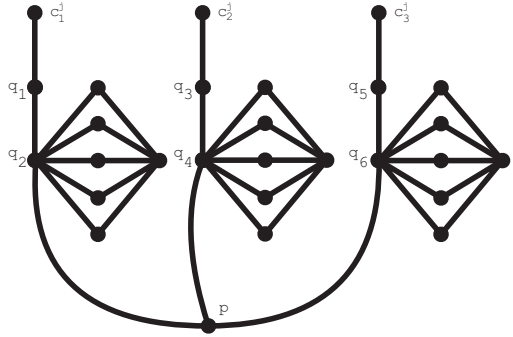}
		\caption{The gadget $\mathcal{L}_{c_j}$. } \label{F11}
	\end{center}
\end{figure}

{\bf Property 14.} Let $G$ be graph such that $\overrightarrow{\chi}_s (G)=2 $ and it has the clause  gadget $\mathcal{S}_{c_j}$ (which is shown in Fig. \ref{F12}) as an induced subgraph.
Also, let $(D,w)$ be an optimal semi-proper orientation of $G$. Then  $   \{S_{(D,w)}(c_1^j),S_{(D,w)}(c_2^j) ,S_{(D,w)}(c_3^j)\}  = \{1,2,2\} $. 
\\ \\
{\bf Proof of Property 14.} By Properties 12 and 13, the proof is clear. $\blacksquare$

Now, we present the construction of $G$ and prove the reduction.\\
\underline{Construction of $G$.} 
We construct the planar bipartite graph $G$ from $ \varphi $ in two steps. \\
{\bf Step 1.} For each variable $x_i \in X$ put a copy of the variable gadget $\mathcal{H}_{x_i}$, which is shown in Fig. \ref{F9}.  Also, for  every clause  $c_j \in C$ put a copy of the clause gadget $\mathcal{S}_{c_j}$, which is shown in Fig. \ref{F12}.  \\ 
{\bf Step 2.} For every clause $c_j\in C$ with the variables $x$, $y$, $z$  add the edges $xc^j_1$, $yc^j_2$ and $zc^j_3$.
Call the resultant graph $G$. It is easy to check that $G$ is planar and bipartite. 

\begin{figure}[ht]
	\begin{center}
		\includegraphics[scale=.45]{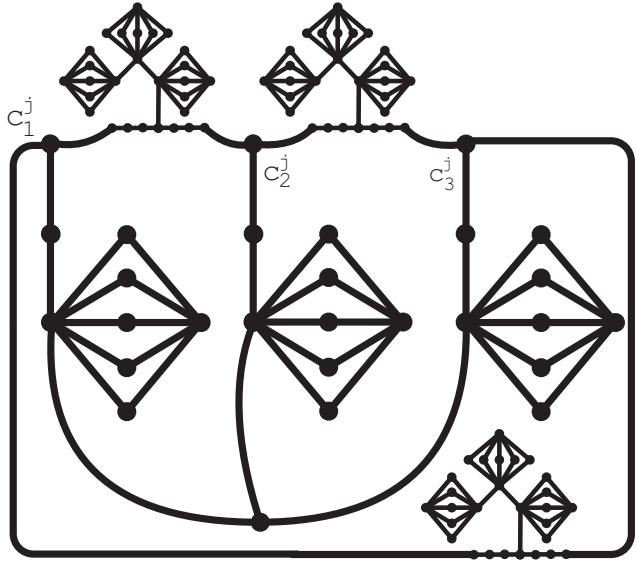}
		\caption{The clause gadget $\mathcal{S}_{c_j}$. } \label{F12}
	\end{center}
\end{figure}

Next, we show that the semi-proper orientation number of the graph $G$ is two   if and only if there is a 1-in-3 satisfying assignment for $ \varphi $.\\
\underline{Proof of $\Rightarrow$}. Let $\overrightarrow{\chi}_s (G)=2 $ and $(D,w)$ be an  optimal semi-proper orientation. Let $\Gamma : X \rightarrow \{{\sf true},{\sf false} \} $ be a function such that if $  S_{(D,w)}(x_i)=2 $, then $\Gamma( x_i)={\sf true}$ and if $  S_{(D,w)} (x_i)=1 $, then $\Gamma( x_i)= {\sf false}$. For each clause $c_j=(x \vee y \vee z)$, by Property 14, we have $$ \{S_{(D,w)}(c_1^j), S_{(D,w)} (c_2^j) , S_{(D,w)} (c_3^j)\} =\{1,2,2\}   .$$ 
So,
$$ \{S_{(D,w)}(x), S_{(D,w)} (y) , S_{(D,w)} (z)\} =\{2,1,1\}   .$$ 
Thus, the function $\Gamma$  is a  1-in-3  satisfying assignment for $ \varphi $. \\
\underline{Proof of $\Leftarrow$}. Assume that $ \varphi $ is satisfiable with the 1-in-3  satisfying
assignment $\Gamma : X \rightarrow \{{\sf true},{\sf false} \} $. For each variable gadget $\mathcal{H}_{x_i}$ orient its edges such that $  S_{(D,w)}(x_i)=2 $ if and only if $\Gamma( x_i)={\sf true}$. Next, for each edge $e$ that connects a vertex of a clause gadget  to a vertex of a variable gadget, put $w(e)=1$ and  orient $e$ from   the vertex of the variable gadget  to  the vertex of the  clause gadget. Finally we need orient the clause gadgets.
Note that for each clause gadget $\mathcal{S}_{c_j}$, we have $ \{S_{(D,w)}(c_1^j), S_{(D,w)} (c_2^j) , S_{(D,w)} (c_3^j)\} =\{1,2,2\}   $. Thus, we can use the orientation which is presented in Fig. \ref{F13} and Fig. \ref{F14} to find an optimal semi-proper orientation for clause gadgets.  This completes the proof of Part (1). 
\\ \\
(2) Note that in the proof of Part (1), we never assign weight one to each edge. Thus, the proof of that part shows the NP-hardness of Part (2) and completes the proof.
}\end{aliii}

\begin{figure}[ht]
	\begin{center}
		\includegraphics[scale=.45]{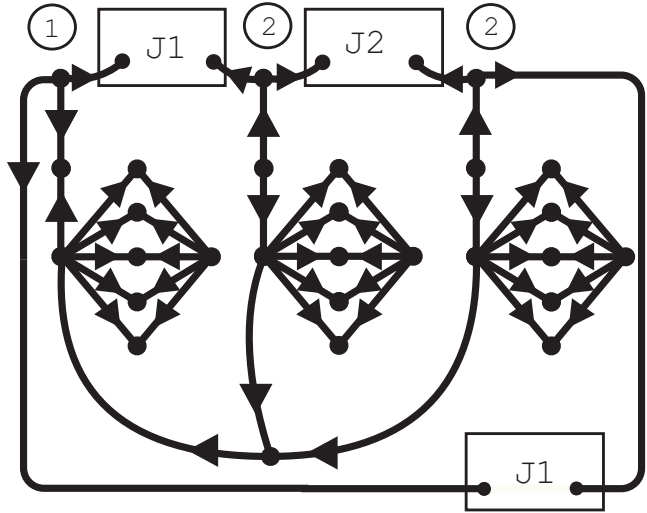}
		\caption{An optimal semi-proper orientation for the gadget $\mathcal{S}_{c_j}$, where the weight of each edge is one and the orientations of the subgraphs $J1$ and $J2$ are presented in Fig. \ref{F14}.} \label{F13}
	\end{center}
\end{figure}

\begin{figure}[ht]
	\begin{center}
		\includegraphics[scale=.65]{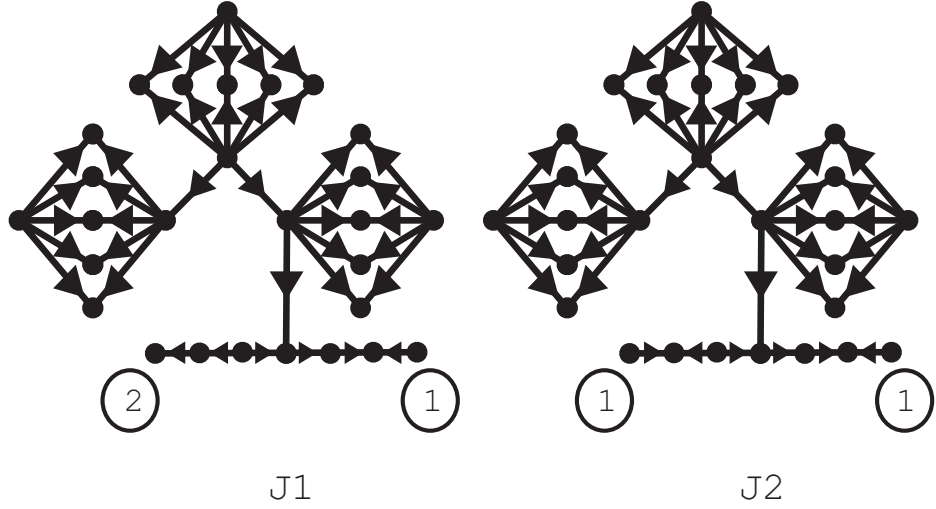}
		\caption{An optimal semi-proper orientation for the gadgets  $J1$ and $J2$, where the weight of each edge is one.} \label{F14}
	\end{center}
\end{figure}

\section{Conclusions and future research}\label{S5}

$\bullet$ In this work, we introduced the notation of semi-proper orientations of graphs and studied some properties of these orientations. It is easy to see that $ \overrightarrow{\chi}_s (G) \leq  \overrightarrow{\chi} (G) $. There are several questions regarding the relationship between the proper orientation number and the semi-proper orientation number of graphs. We pose some of them here.

\begin{prob}
	Is there any constant number $c$ such that $   \overrightarrow{\chi} (G)-  \overrightarrow{\chi}_s (G) \leq c $?
\end{prob}

\begin{prob}
	Is there any important family of graphs such that for  graphs in that family there is a polynomial time algorithm to compute  $\overrightarrow{\chi}_s (G)$ but computing $\overrightarrow{\chi} (G)$ is NP-hard?
\end{prob}

$\bullet$ The problem of determining an upper  bound for the  proper orientation numbers of planar graphs is a well-known open problem in this area \cite{MR3293286, araujo2018weighted}. We pose the same problem.

\begin{prob}
	Does there exist a constant number $c$ such that $    \overrightarrow{\chi}_s (G) \leq c $  for all planar graphs $G$?
\end{prob}

$\bullet$  We proved that every graph $G$ has an optimal semi-proper orientation $(D,w)$ such that the weight of each edge is one or two. Also, we proved that determining whether a given planar graph $G$ with $\overrightarrow{\chi}_s (G)=2 $ has an optimal semi-proper orientation $(D,w)$ such that the weight of each edge is one is NP-complete. It is interesting to find some families of graphs such that each graph in those families has an optimal semi-proper orientation $(D,w)$ such that the weight of each edge is one. We pose the following problem.

\begin{prob}
	Is there any polynomial time algorithm to determine whether a given bipartite graph $G$ has  an optimal semi-proper orientation $(D,w)$ such that the weight of each vertex is one?
\end{prob}

$\bullet$  Furthermore, in this work we proved that the problem of determining the semi-proper orientation number  of planar bipartite graphs is NP-hard. Regarding the complexity of computing the proper orientation number of regular graphs it was shown in \cite{MR3095464} that it is NP-complete to decide whether
the proper orientation number of a given 4-regular graph
is 3. What can we say about the complexity of computing the semi-proper orientation number of regular graphs?

\begin{prob}
Determine the computational complexity of computing the semi-proper orientation number of regular graphs.
\end{prob}

\bibliographystyle{plain}
\bibliography{Semi-proper}

\end{document}